\documentclass[conference]{IEEEtran}

\IEEEoverridecommandlockouts

\usepackage{cite}
\usepackage{amsmath,amssymb,amsfonts}
\usepackage{algorithmic}
\usepackage{graphicx}

\usepackage{textcomp}
\usepackage{xcolor}
\usepackage{hyperref}
\usepackage[capitalise]{cleveref} 
\usepackage{makecell} 
\usepackage{bm} 

\def\BibTeX{{\rm B\kern-.05em{\sc i\kern-.025em b}\kern-.08em
    T\kern-.1667em\lower.7ex\hbox{E}\kern-.125emX}}

\usepackage{mathtools}

\DeclareMathOperator{\IGate}{\mathsf{I}} 
\DeclareMathOperator{\XGate}{\mathsf{X}}   
\DeclareMathOperator{\ZGate}{\mathsf{Z}}   
\DeclareMathOperator{\PGate}{\mathsf{P}}  
\DeclareMathOperator{\HGate}{\mathsf{H}}   
\DeclareMathOperator{\CNOTGate}{\mathsf{CNOT}}   
\DeclareMathOperator{\TGate}{\mathsf{T}}

\newcommand{\sk}{{\it sk}}
\newcommand{\pk}{{\it pk}}

\newcommand{\SWAP}{{\mathsf{SWAP}}}
\def\ket#1{{\lvert}#1\rangle}

\let\code\mathtt

\usepackage{tikz}
\usepackage{euscript}

\begin{document}

\title{Demonstrating Quantum Homomorphic Encryption Through Simulation
 \thanks{We acknowledge funding support under the Digital Horizon Europe project FoQaCiA, Foundations of quantum computational advantage, grant no.~101070558, together with the support of the Natural Sciences and Engineering Research Council of Canada (NSERC), and the University of Ottawa’s Research Chairs program.}
}

\author{\IEEEauthorblockN{Sohrab Ganjian}
\IEEEauthorblockA{\textit{Department of}\\ \textit{Mathematics \& Statistics} \\
\textit{University of Ottawa}\\
Ottawa, ON, Canada \\
sohrab.g@uottawa.ca}
\and
\IEEEauthorblockN{Connor Paddock}
\IEEEauthorblockA{\textit{Department of}\\ \textit{Mathematics \& Statistics} \\
\textit{University of Ottawa}\\
Ottawa, ON, Canada \\
cpaulpad@uottawa.ca}
\and
\IEEEauthorblockN{Anne Broadbent}
\IEEEauthorblockA{\textit{Department of}\\ \textit{Mathematics \& Statistics} \\
\textit{University of Ottawa}\\
Ottawa, ON, Canada \\
abroadbe@uottawa.ca}
}

\maketitle

\begin{abstract} Quantum homomorphic encryption (QHE), allows a quantum cloud server to compute on private data as uploaded by a client. 
We provide a proof-of-concept software simulation for QHE, 
according to the ``EPR'' scheme of Broadbent and Jeffery, for universal quantum circuits. We demonstrate the near-term viability of this scheme and provide verification that the additional cost of homomorphic circuit evaluation is minor when compared to the simulation cost of the quantum operations. Our simulation toolkit is an open-source Python implementation, 
that serves as a step towards further hardware applications of quantum homomorphic encryption between networked quantum devices.

\end{abstract}


\section{Introduction}\label{sec:introduction}

Cloud computation allows clients with limited computational resources to offload computations to more powerful remote servers. This framework is very convenient in an age where transmitting data is quite economical while computational resources remain comparatively expensive.  However, this paradigm brings a new concern, which is the 
 privacy of the client's data. The breakthrough work of \cite{Gen09} established the first fully homomorphic encryption (HE) scheme. An HE scheme allows a server to run any computation on a client's encrypted data, without ever needing to decrypt the data. HE schemes are integral to several cloud computing protocols, such as delegated computing, two-party secure computation, and zero-knowledge proofs \cite{BB12}.

The advent of quantum computing delivers an alternative perspective to computing based on the physics of quantum mechanics and holds the promise of substantial benefits \cite{Sho99,Gro96}. However, the technology's enormous cost and complex engineering requirements mean that for most users, the technology will likely remain inaccessible, even in the long term. 
Nevertheless, quantum computing resources are already available through the cloud model from several companies \cite{MSK19}, enabling clients to upload classical data to remote servers for quantum computation.

Quantum homomorphic encryption (QHE) aims to achieve the objectives of HE with quantum data and quantum circuits. There have been numerous contributions in the theory of quantum homomorphic encryption \cite{BJ15,TKO+16, ADSS17,OTF18, Lia20}, with the breakthrough work of \cite{Mah18b} providing a theoretical construction of QHE with a fully classical client.

In this work, we consider a scenario where a client has access to a quantum ``encryption/decryption device'', which allows them to encrypt, transmit, receive, and decrypt quantum states, but not perform universal quantum computation. Additionally, we focus on implementing the ``EPR'' QHE scheme introduced in \cite{BJ15}, due to the modular nature of its security and construction. Indeed, the EPR scheme primarily leverages its security from an underlying classical HE scheme, which can be readily integrated into the QHE scheme.
Moreover, the technological requirements and cryptographic techniques required in the EPR scheme are more straightforward in comparison to those in \cite{Mah18, DSS16}, as the scheme does not rely on trapdoor-claw-free functions or polynomially many additional qubits in the security parameter of the classical HE scheme. Such requirements are currently out of reach (practically in the former case, and physically in the latter case).\looseness=-1

This work provides a software demonstration of the EPR QHE scheme in Python. Our preliminary testing affirms that the computational cost of the classical HE is negligible relative to the cost of the quantum simulation. Our software implementation provides further evidence that the EPR scheme is a good candidate for physical applications using networked quantum devices, as already demonstrated by experiments such as \cite{TFB+20}. In addition, several subsequent QHE schemes share common elements and primitives developed in the EPR scheme. With this in mind, we see our software implementation as an opportunity for the community to leverage our tools for subsequent implementations of QHE.

To the best of our knowledge, our implementation is the first open-source toolkit for simulating quantum homomorphic encryption. Our software implementation relies heavily on the NumPy library \cite{HMJ+20}. For the HE component of our implementation, we selected a Python implementation of the Brakerski-Fan-Vercauteren (BFV) levelled classical homomorphic encryption scheme in \cite{Era20}, that is based on the works \cite{FV12, Bra12}. The security of this classical HE scheme is based on the hardness of the learning-with-errors (LWE) problem, which is reputed to be post-quantum secure~\cite{Reg05}.

The remainder of this work is as follows: \cref{sec:prelims} briefly outlines the EPR scheme, \cref{sec:implementation} provides the details of our implementation, design choices, and insight, \cref{sec:test_cases} includes the parameters of our implementation, as well as some data and analysis of the simulation, and finally, \cref{sec:future_work} discusses future work and questions left open by this work.

\section{Overview of QHE Scheme}\label{sec:prelims}

In \cite{BJ15}, Broadbent and Jeffrey proposed a QHE scheme based on the quantum one-time pad (QOTP) encryption alongside a classical HE scheme. Their ``EPR scheme'' is homomorphic for the universal Clifford+$\mathsf{T}$ gate set, and the security of the scheme relies on the computational security of the classical HE scheme.

We briefly review the EPR scheme. For background on quantum computing, we refer the reader to \cite{NC10}. Each qubit $\ket{\psi}$ is encrypted using the QOTP and a pair of secret key bits $(a,b)$. The QOTP maps each qubit to the state $\XGate^{a} \ZGate^{b} \ket{\psi}$, where $\XGate$ and $\ZGate$ are the standard Pauli matrices.
Using a classical HE scheme, the client encrypts the one-time-pad keys and provides the encrypted keys to the server along with the encrypted qubits.

Given the encrypted data, the server performs homomorphic evaluation of a quantum circuit using the following technique. Recall that any quantum circuit on $n$-qubits can be decomposed into a circuit consisting of one- and two-qubit gates from the set $\mathfrak{C}\cup \{\TGate\}$, where $\mathfrak{C}=\{\XGate,\ZGate,\HGate,\PGate,\CNOTGate\}$ is the set of  Clifford gates \cite{BMP+00, NC10}. Here $\HGate$ is the Hadamard gate, $\CNOTGate$ is the controlled-NOT gate, which applies an~$\XGate$ on the second register of a two-qubit system only if the first qubit is in the state $\ket{1}$. The $\PGate$ and $\TGate$ gates are defined by matrices $\begin{bmatrix}
    1 & 0 \\
    0 & i
\end{bmatrix}$ and $\begin{bmatrix}
    1 & 0 \\
    0 & e^{\frac{i\pi}{4}}
\end{bmatrix}$, respectively. It follows that if a quantum server can execute each of these gates homomorphically, the server can perform homomorphic evaluation of any quantum circuit. With this in mind, the scheme EPR consists of two components: one for evaluating Clifford gates homomorphically and the other for evaluating non-Clifford ($\TGate$) gates.

To address the Clifford gates, the scheme leverages the structure of the Clifford group and the Pauli gates used in the QOTP. Specifically, for any Pauli $\mathsf{Q}$ and Clifford operation~$\mathsf{C}$, there exists a Pauli $\mathsf{Q}'$ such that $\mathsf{C}\mathsf{Q} = \mathsf{Q}'\mathsf{C}$. When $\mathsf{Q}=\XGate^a\ZGate^b$ is the Pauli operation in the QOTP with key $(a,b)$, application of a Clifford gate $\mathsf{C}$ results in the updated Pauli $\mathsf{Q}'=\XGate^{a'}\ZGate^{b'}$. Hence, for the server to perform a Clifford gate homomorphically it can perform the gate and update the keys accordingly. The key update rules $(a,b)\mapsto (a',b')$ for each Clifford gate are described in \cite{BJ15,Bro15} and for convenience are listed in \cref{tbl:key update rules} Since the key updates for the Clifford gates are classical operations, the server can perform these key updates using the classical HE scheme\footnote{We remark that a classical additive homomorphic encryption scheme would suffice for the Clifford key updates.} in which  $a$ and $b$ are provided in an encrypted form only. 

Evaluating the non-Clifford $\TGate$ gate is less straightforward. Since the  $\TGate$ gate is not Clifford, when it is applied to the Pauli $\XGate^a \ZGate^b$, the result picks up an additional phase gate depending on the value of $a$. That is, $
\TGate \XGate^a \ZGate^b = \XGate^a \ZGate^{a \oplus b} \PGate^a \TGate
$. To account for this non-Clifford $\PGate$ gate, the scheme makes use of a gate-teleportation gadget, called the ``$\TGate$-gate gadget''. This gadget is shown in Figure 2. of \cite{BJ15}. We provide a modified version of the gadget in \cref{fig:T-gadget-EPR}. The $\TGate$-gate gadget requires that an entangled (EPR) pair $\ket{\Phi} = \frac{1}{\sqrt{2}}(\ket{00}+\ket{11})$ be shared between the client and server. This gadget effectively enables a Pauli key update rule $a\mapsto a \oplus c$ and $b\mapsto a \oplus b \oplus c\cdot a \oplus k$ for $\TGate$ gate evaluation by the server. However, because $k$ is unknown to the server, a portion of the key update computation is done symbolically. Again, this $\TGate$ gate key update computation is performed on the encrypted keys through the classical HE scheme.

After the circuit is completely evaluated, the server returns the quantum state and the modified keys to the client. The client decrypts the keys and recovers the quantum state by inverting the QOTP. Here we remark that the decryption algorithm for the EPR scheme depends on the gates and measurements required by the $\TGate$-gate gadget as seen in \cref{fig:T-gadget-EPR}. Specifically, if $R$ is the number of $\TGate$ gates in the circuit evaluated by the server, the runtime of the decryption is $O(R^2)$. The dependence of the decryption algorithm on the circuit means that the EPR scheme is not compact\footnote{Compactness is a property of HE (and QHE) which insists the decryption run time is independent of the size of the circuit evaluated by the server. Without compactness, one can imagine a trivial scheme where the server appends the circuit to the ciphertext and demands the client perform the computation as part of the decryption.} We refer the reader to \cite{BJ15} for a detailed analysis of the correctness and security guarantees of the scheme as well as for information on how it achieves a certain degree of compactness.

\section{Implementation Details and Insights}\label{sec:implementation}

In this section, we present the details of the EPR scheme implementation, available online~\cite{Gan24}. We emphasize the design choices made in the development process as well as the unique challenges we encountered in lifting the theoretical description of the protocol into software. 

\subsection{Implementation Details}\label{sub:details}
We now provide a detailed description of our implementation which follows closely the documentation in our code~\cite{Gan24}.

    \textbf{Key Generation} --- Our implementation uses the key generation functions $\code{BFVKeyGenerator}$ from the Python BFV scheme implemented in \cite{Era20}. Using this package, we generate a public key $\pk$ and a secret key $\sk$ to be used in the encryption and decryption of the classical one-time-pad keys.

    \textbf{Encryption} --- Given an input $n$-qubit quantum state $\ket{\psi}$ and a pair of randomly generated $n$-bit keys $(a,b)$, the client uses the public key $\pk$ to produce the encryption pair $(\Tilde{a}, \Tilde{b})$ along with an encrypted state $\Tilde{\ket{\psi}}$, where the state $\Tilde{\ket{\psi}}$ is obtained from the quantum one-time-pad, which takes as input the pair~$(a,b)$. 

\begin{figure}[tb]
    \includegraphics[scale=0.9]{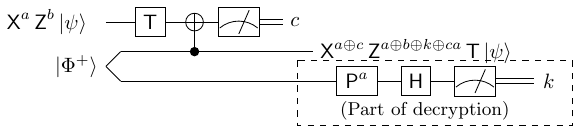}
    \caption{\label{fig:T-gadget-EPR}Evaluation protocol for an application a $\TGate$ gate. The dashed box shows part of the decryption procedure, which happens at some point in the future after the complete evaluation is finished.}
\end{figure}

    \textbf{Homomorphic Evaluation} --- The server's description of the quantum circuit is represented by $\mathcal{C}$, a list of strings in the alphabet $\{ \IGate, \XGate, \ZGate, \HGate, \PGate, \CNOTGate, \TGate \}$. Each element of the list corresponds to a layer of the quantum circuit and each symbol in the string represents the gate applied to qubit(s) $i$ in the $j$th layer, namely $\mathcal{C}^{(i)}_{j}$, where $1\leq j\leq \ell$ ranges through the $\ell$ layers of the circuit, and $1\leq i\leq n$ for the $n$-qubits. Before evaluating the circuit, the server appends $R$ EPR pairs to the input of the computation, where $R$ is the total number of $\TGate$ in $\mathcal{C}$. Once completed, the server applies a series of $\SWAP$ gates so that all the wires associated with EPR pairs used for computing any $\mathsf{T}$ gates appear in the required order (see \cref{fig:T-gadget-EPR}). The server applies the circuit $\mathcal{C}$ layer by layer. Each layer $\mathcal{C}_j$ is implemented by a unitary $\mathcal{L}_j$ on $n+2R$-qubits, consisting of at most $n$ non-identity gates.
    
    The server implements $\mathcal{L}_j$ gate by gate, inserting identity gates where required. If the gate $\mathcal{C}^{(i)}_{j}$ is Clifford, then the keys $\left(\Tilde{a}^{(i)}_j, \Tilde{b}^{(i)}_{j} \right)$ are updated according to \cref{tbl:key update rules}. On the other hand, if the gate $\mathcal{C}^{(i)}_{j}$ is a $\TGate$ gate, the server stores the ciphertext $\Tilde{a}^{(i)}_{j}$ in a list $\Tilde{\mathcal{P}}$, and applies the $\TGate$ gate followed by $\CNOTGate_{2 \rightarrow 1}$ and measures the wire that initially held the encrypted state obtaining the bit $c^{(i)}_j$ (see \cref{fig:T-gadget-EPR}). Storage of the ciphertext in the list $\Tilde{\mathcal{P}}$ will be used as part of the decryption. The server updates the wire labels holding the evaluated qubit as the encrypted qubit has been ``teleported'' to the 1st wire of the EPR pair.
    
    To make the homomorphic update for the $\TGate$, the measured bit $c^{(i)}_j$ is encrypted using the BFV scheme public key $\pk$. The resulting ciphertext $\Tilde{c}^{(i)}_{j}$ is stored in a list $\Tilde{\mathcal{M}}$. Now, the server performs the homomorphic evaluation by computing $\Tilde{a}^{(i)}_{j-1} \mapsto \Tilde{a}^{(i)}_{j-1} \oplus \Tilde{c}^{(i)}_{j}$ and $\Tilde{b}^{(i)}_{j-1} \mapsto \Tilde{a}^{(i)}_{j-1} \oplus \left(\Tilde{a}^{(i)}_{j-1} \cdot \Tilde{c}_{j}^{(i)}\right) \oplus \Tilde{b}^{(i)}_{j-1}$, which accounts for the portion of the homomorphic $\TGate$ gate key update with \emph{known} variables. To account for the \emph{unknown} variables in the homomorphic $\TGate$ gate key update, the server also performs a symbolic computation tracking the remaining unknown values $k_{j}^{(i)}$. Here the encrypted keys are treated as unknown symbolic variables. Specifically, for a $\TGate$ gate,  $\hat{a}^{(i)}_{j-1} \mapsto \hat{a}^{(i)}_{j-1} \oplus \hat{c}^{(i)}_{j}$ as usual. However, $\hat{b}^{(i)}_{j-1} \mapsto \hat{a}^{(i)}_{j-1} \oplus \left(\hat{a}^{(i)}_{j-1} \cdot \hat{c}_{j}^{(i)}\right) \oplus \hat{b}^{(i)}_{j-1} \oplus \bm{\hat{k}_{j}^{(i)}}$. We use the notation $\hat{(\cdot)}$ to denote symbolic versions of the variables. 
    We need to perform symbolic updates for the Clifford gates as well, according to \cref{tbl:key update rules} as expected; however, as we discuss later, the unknown values of $k_{j}^{(i)}$ can propagate into the keys updates of subsequent layers of Clifford key updates.
    
    In a list $\mathcal{S}$, the server stores the symbolic value of $\hat{a}$ and $\hat{b}$ at the end of the circuit. The list also includes a record of each value of $\hat{a}$ before the application of a $\TGate$ gate. Likewise, the order in which $\TGate$ gates appear in the circuit is transcribed in a list~$\mathcal{T}$. Both lists are used in the decryption.

    \textbf{Decryption} --- Using the secret key $\sk$, the client decrypts the ciphertexts $(\Tilde{a}, \Tilde{b})$ as well as the encrypted bits in $\Tilde{\mathcal{M}}$ and~$\Tilde{\mathcal{P}}$. We denote these decrypted lists as $\mathcal{M}$ and $\mathcal{P}$, respectively. The client begins decrypting by the order specified in the list $\mathcal{T}$. If $\mathcal{T}$ is empty, the client can recover the evaluated quantum state by applying the quantum one-time pad using the decrypted keys $(a,b)$. Otherwise, decryption proceeds as follows: the first conditional phase gate exponent does not depend on any unmeasured values. Therefore, in this case the client can perform the conditional $\PGate^{a^{(i)}_j}$ by reading the value of $a^{(i)}_{j}$ from the decrypted values of $\mathcal{P}^{(i)}_j$. The client applies the $\HGate$ gate and then proceeds to measure $k^{(i)}_{j}$, storing this value in a list $\mathcal{K}$ (see \cref{fig:T-gadget-EPR}).
    
    For the decryption of any subsequent $\TGate$ gates, the client may need to apply corrections to the values in $\mathcal{P}$, since those values could depend on yet-to-be-measured values of $k$, which were unknown during the evaluation. To apply theses corrections, the client looks up the $\mathcal{S}^{(i)}_j$ entry holding the symbolic expression of the key updates. The client substitutes $0$ for expressions involving $\hat{a}$ and $\hat{b}$ and the corresponding values of $\hat{c}$ and $\hat{k}$ from the appropriate entries of lists $\mathcal{M}$ and~$\mathcal{K}$ to evaluate the polynomial, which outputs the bit necessary to correct the exponent value of the $\PGate$ gate. Only after this correction, can the client apply the decryption part of the $\TGate$-gate gadget. The new measured $k^{(i)}_{j}$ value is then appended in the $\mathcal{K}$ list. Upon completing this process for all the $\TGate$ gates, the client repeats the classical bit correction procedure to obtain the decrypted $a$ and $b$ values. Lastly, the client traces out all the unnecessary wires. It is only now that the client can apply the quantum one-time-pad to obtain the decrypted quantum state.\looseness=-1

\begin{table}[tb]
\caption{Key Updates after applying a Clifford Gate}
\centering
\begin{tabular}{|c|c|c|}
\hline
Gate & Before  & After            \\
\hline
$\IGate, \XGate, \ZGate$  & $(a,b)$ & $(a,b)$          \\
\hline
$\HGate$  & $(a,b)$ & $(b,a)$          \\
\hline
$\PGate$  & $(a,b)$ & $(a,a \oplus b)$ \\
\hline
$\CNOTGate$ & 
\makecell{$(a^{(i)},b^{(i)}),$ \\ $(a^{(i+1)},b^{(i+1)})$} & 
\makecell{$(a^{(i)}, b^{(i)} \oplus b^{(i+1)}),$ \\ $(a^{(i)} \oplus a^{(i+1)}, b^{(i+1)})$} \\
\hline
\end{tabular}
\label{tbl:key update rules}
\end{table}

\subsection{Design Choices and Insights}

Here we provide some additional discussion of the implementation. We highlight some of the more challenging components, as well as provide insight into some of the more complex parts of the software design.

\textbf{Symbolic Computations} --- A challenging component of the implementation was carrying out the conditional phase gate used in the $\TGate$-gate gadget. At first glance, performing $\PGate^a$ may seem straightforward as only the $b$ component of the homomorphic key update for $\TGate$ appears to contain any yet-to-be-measured bits $k$. However, this is not the case, as unmeasured bits can appear in the $a$ component of the key. Explicitly, decryption following homomorphic evaluation of the circuit $\TGate \HGate \TGate$ would result in a $k$ value in the~$a$ component before the application of the second $\TGate$, since the $\HGate$ gate swaps the components $a$ and $b$. Similarly, one might think that the key components of an encrypted qubit on a wire containing no only Clifford gates will not contain these unmeasured values of~$k$. However, this is not necessarily the case either. For instance, the circuit $\CNOTGate \times (\IGate \otimes \TGate)$ will result in unmeasured $k$ values to be ``spilled over" to the $b$ component of the first qubit at the end of the evaluation, despite no $\TGate$ being applied on the first qubit. These examples illustrate the important role of symbolic computation for keeping track of the values in the key updates. In our case, this was achieved through the use of the SymPy package \cite{ACP+17}.

\textbf{Wire Relabelling} --- The original EPR scheme in \cite{BJ15} assumes that the appended Bell pairs appear adjacent to the corresponding qubit(s). However, in practice one needs to account for the relabelling of wires. We achieve this by performing $\SWAP$ gates to place the EPR pairs adjacent to the corresponding qubit as in \cref{fig:T-gadget-EPR}. Although we perform $\SWAP$ operations as needed, one could alternatively track these within the implementation and synthesize the required permutation of the output registers before decryption. In either case, one could be concerned that the $\SWAP$ gates need to be performed homomorphically. Fortunately, $\SWAP$ is  Clifford and the key updates match our intuition. That is, the  keys $(a_1, b_1),(a_2,b_2)$ get mapped to $(a_2,b_2), (a_1,b_1)$ after a $\SWAP$ is performed.

\textbf{Generalized Controlled-NOTs} --- In the software implementation, we develop tools for tracking the wire(s) where gates were applied. This is a practical way to account for the fact that the $\TGate$-gate gadget teleports the encrypted state to another wire. Not only is this tracking of information crucial in the evaluation and decryption procedure but also in the case where multiple $\TGate$ gates are applied to the same qubit. In particular, when the $\TGate$-gate gadget requires a $\CNOTGate$ gate between non-adjacent wires. This is not possible with the standard $\CNOTGate_{2 \rightarrow 1}$. To address this, we constructed a function $\code{controlled\_not\_constructor  (control, target,}$ \newline $\code{number\_of\_qubits)}$ for performing $\CNOTGate$s between non-adjacent wires. In the case $\code{target} < \code{control}$, the function creates the following $n$-qubit gate

\begin{align}
\begin{split}
 \mathsf{CX}_{c, t, n} &= \left[ \IGate^{\otimes(c-1)} \otimes | 0 \rangle \langle 0 | \otimes \IGate^{\otimes(n-c)} \right] \\
 &+ \left[ \IGate^{\otimes(t-1)} \otimes \XGate \otimes \IGate^{\otimes(c-t-1)} \otimes | 1 \rangle \langle 1 | \otimes \IGate^{\otimes (n-c)} \right],
 \end{split}
\end{align}
where we use the convention that $\IGate^{\otimes 0} = 1$.

\section{Test Cases and Analysis}\label{sec:test_cases}

In this section, we provide details of our test implementations and some analysis of our initial findings. Our simulations were conducted on an Intel core i9-10885H with 32GB RAM using Python 3.8. The security parameters for the BFV scheme include a polynomial ring of degree~$8$, a plaintext modulus of~$17$, and a ciphertext modulus of $8 \times 10^{12}$, as used by the example provided in \cite{Era20}.

\begin{figure}[htbp]
    \centering

    \begin{minipage}[t]{0.5\textwidth}
        \centering
        \includegraphics[width=\linewidth]{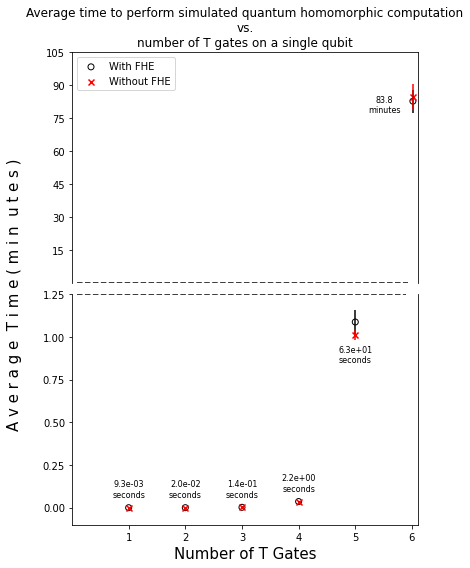}
        \caption{Runtime of the simulation of quantum homomorphic encryption based on the number of $\TGate$ gates with and without classical homomorphic encryption. The displayed runtime is the average of the two schemes.}
        \label{fig:NumberOfTGateVTime}
    \end{minipage}
    \hfill  
    \begin{minipage}[t]{0.5\textwidth}  
        \centering
        \includegraphics[width=\linewidth]{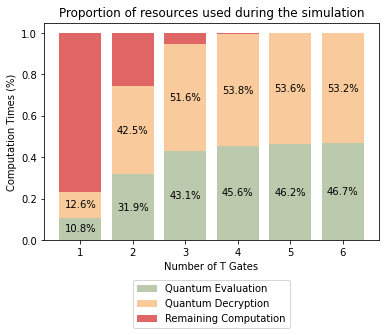}
        \caption{Proportion of computational resources used during the simulation. Remaining Computation includes any purely classical computations during the protocol, including evaluation and decryption, as well as the quantum one-time pad operation during the encryption.}
        \label{fig:ComputationProportions}
    \end{minipage}

\end{figure}

\cref{fig:NumberOfTGateVTime} displays the average runtimes of the simulated EPR scheme for a circuit consisting of $R\in \{1,\ldots,6\}$ $\TGate$ gates on a single qubit.
 The test aims to establish a benchmark for the cost of simulating the quantum operations by quantifying the time taken by the simulation of the EPR scheme with and without the homomorphic key updates. In the test runs without the encrypted key updates\footnote{Details of the EPR scheme with unencrypted keys are included in our software library \cite{Gan24}, including the implementation that was used to collect the data in \cref{fig:NumberOfTGateVTime}.}, we assume the circuit is known to the client and all the updates are performed by the client during the decryption. We stress that this is not the case in the EPR scheme with the encrypted key updates outlined in \cref{sub:details}. The data in \cref{fig:NumberOfTGateVTime} from the circuits with $1$ to $5$ $\TGate$ gates are averages from a $10$ sample runs, whereas in the $6$ $\TGate$ gate circuit the average was only taken from $3$ sample runs due to computational limitations. The error bars in \cref{fig:NumberOfTGateVTime} represent a standard deviation.

Recall that if the $n$-qubit quantum circuit has $R$ $\TGate$ gates, the scheme requires an additional $2R$ qubits. This implies that the simulation amounts to performing quantum operations on an $n+2R$-qubit system, \emph{i.e.}~multiplying $2^{n+2R}\times 2^{n+2R}$ matrices, which requires an exponential number of operations in the parameters $n$ and $R$. The exponential nature of operations manifests itself in \cref{fig:NumberOfTGateVTime} by noting the marked increase in the computational time between the runs with $5$ and $6$ $\TGate$ gates, from around 1 minute to almost 1 hour and a half, respectively.

On its own, the data presented in \cref{fig:NumberOfTGateVTime} is too coarse to capture the relationship between the cost of the classical and quantum operations. More specifically, we want to better understand how the classical components of the code, such as homomorphic key updates, wire tracking and symbolic computations contribute to the overall runtime of the simulation. \cref{fig:ComputationProportions} displays the proportion of computational resources used with the classical homomorphic encryption. The data in \cref{fig:ComputationProportions} affirms our suspicion that the classical components of the code do not have much impact on the runtime of the simulation. As expected \cref{fig:ComputationProportions} shows that as the number of $\TGate$ gates increase, a larger portion of the runtime is used in simulating the quantum operations.

Interestingly, \cref{fig:ComputationProportions} shows that the decryption takes slightly more computational resources than the evaluation. A plausible explanation is by noting that for every $\TGate$ gate in the circuit, there are two gates applied during evaluation: a $\TGate$ followed by a $\CNOTGate$, and two more in the decryption: a conditional $\PGate$ and an $\HGate$. Additionally, in the decryption, the client needs to perform the quantum operation to trace out unnecessary wires. From the data in \cref{fig:NumberOfTGateVTime} for the EPR scheme with encrypted keys, and the data in \cref{fig:ComputationProportions}, we can deduce that the runtime of the decryption grows as the number of $\TGate$ gates increases, providing evidence that the scheme is not compact.

The attentive reader will note that in the case of the 6 $\TGate$ gate experiment in \cref{fig:NumberOfTGateVTime}, on average the EPR scheme without encrypted keys took slightly longer than the EPR scheme with encryption. This is surprising since one expects the addition of HE to require more resources. Nonetheless, this anomaly can be explained by the fact that the classical operations use a negligible amount of time in comparison to the quantum operations as evident in \cref{fig:ComputationProportions}. This observation, along with the overlap of the error bars from the 5 and 6 $\TGate$ gate experiments (with and without encrypted keys) supports this claim. Variability of the runtime in each test can be affected by factors such as background tasks and CPU variability.

\section{Conclusion and Future Work}\label{sec:future_work}

In this work, we developed a Python implementation capable of simulating QHE. In particular, our software implementation of the EPR scheme from \cite{BJ15} works for universal quantum circuits. We discussed the details and challenges of bridging the theoretical details of the scheme into practice. We provided some preliminary data which suggests that the cost of classical homomorphic encryption is nominal relative to the quantum simulation costs. We hope that the quantum community and enthusiasts will benefit from our software library, towards advancing the frontiers of quantum cloud computing. 

There are several developments that we believe could improve the performance of our simulation toolkit. The first of which could be improving the performance of the quantum operations. One method could be synthesizing the unitary circuit for an entire layer, so that all the gates in a layer are performed simultaneously, rather than one gate at a time. Another way would be to reduce the dimension of the quantum operations by tracing out the measured ``extra'' wires during the evaluation of the $\TGate$-gate gadget. Doing this could reduce the wires from $n+2R$ to just $n+R$ at the end of the evaluation, saving a factor of $2^R$ in the decryption runtime. Further improvements are possible in other areas. 
For example, the current implementation requires the quantum circuit description in a certain form. In particular, the circuit must be synthesized so only $\CNOTGate$ gates act on adjacent qubits. An improvement to the implementation would involve handling arbitrary controlled-NOT gates where the control and target qubits may be in any position.

A next step would be to expand our library using techniques from follow-up works to the EPR scheme. For instance, in \cite{DSS16} the EPR scheme was improved by introducing an alternative $\TGate$-gate gadget which results in a compact levelled scheme for performing non-Clifford gates homomorphically. Implementing this new gadget would allow the simulation of QHE on quantum circuits with a polynomial number of $\TGate$ gates.
Verifiability is an important ingredient for secure delegated quantum computing. The addition of a verification procedure is a desirable feature, as often the client may want to delegate a quantum circuit to the server while ensuring that the output is correct. In \cite{ADSS17} the QHE scheme of \cite{DSS16} was made verifiable. Adding a verification component to our implementation would be an interesting next step. Finally, the EPR scheme leaks information about the quantum circuit to the client. In particular, in the honest case, the client receives a qubit for each $\TGate$ gate to measure in the decryption procedure. This is an issue in the case where the server wants the circuit to remain private. In \cite{DSS16} it is suggested that the server can take measures to maintain circuit privacy, by adding randomization layers to the circuit. Future work towards including such features in our library could provide additional utility to the QHE scheme, particularly in the context of proprietary quantum algorithms.

\bibliographystyle{IEEEtran}
\makeatletter\@ifundefined{url}{\newcommand{\url}[1]{\texttt{#1}}}{}\@ifundefined{href}{\newcommand{\href}[2]{\texttt{#2}}}{}\@ifundefined{mathbb}{\newcommand{\mathbb}[1]{#1}}{}\makeatother

\end{document}